\documentclass[twocolumn]{aastex63}
\usepackage{graphicx}
\usepackage{subfigure}
\usepackage{color, hyperref, epsfig}
\usepackage{apjfonts, natbib}
\usepackage{appendix}
\usepackage{float}
\usepackage{bm}

\newcommand{\lum}{erg\,s\ensuremath{^{-1}}}

\newcommand{\mbh}{\ensuremath{M_\mathrm{BH}}}

\shorttitle{AT~2023clx}
\shortauthors{ZHU et al.}

\begin{document}

\title{\Large AT~2023clx: the Faintest and Closest Optical Tidal Disruption Event Discovered in Nearby Star-forming Galaxy NGC~3799}

\author[0000-0003-3824-9496]{Jiazheng Zhu}
\affiliation{CAS Key laboratory for Research in Galaxies and Cosmology,
Department of Astronomy, University of Science and Technology of China, 
Hefei, 230026, China; jiazheng@mail.ustc.edu.cn, jnac@ustc.edu.cn}
\affiliation{School of Astronomy and Space Sciences,
University of Science and Technology of China, Hefei, 230026, China}

\author[0000-0002-7152-3621]{Ning Jiang}
\affiliation{CAS Key laboratory for Research in Galaxies and Cosmology,
Department of Astronomy, University of Science and Technology of China, 
Hefei, 230026, China; jiazheng@mail.ustc.edu.cn, jnac@ustc.edu.cn}
\affiliation{School of Astronomy and Space Sciences,
University of Science and Technology of China, Hefei, 230026, China}

\author[0000-0002-1517-6792]{Tinggui Wang}
\affiliation{CAS Key laboratory for Research in Galaxies and Cosmology,
Department of Astronomy, University of Science and Technology of China, 
Hefei, 230026, China; jiazheng@mail.ustc.edu.cn, jnac@ustc.edu.cn}
\affiliation{School of Astronomy and Space Sciences,
University of Science and Technology of China, Hefei, 230026, China}

\author[0000-0001-7689-6382]{Shifeng Huang}
\affiliation{CAS Key laboratory for Research in Galaxies and Cosmology,
Department of Astronomy, University of Science and Technology of China, 
Hefei, 230026, China; jiazheng@mail.ustc.edu.cn, jnac@ustc.edu.cn}
\affiliation{School of Astronomy and Space Sciences,
University of Science and Technology of China, Hefei, 230026, China}

\author[0000-0003-4959-1625]{Zheyu Lin}
\affiliation{CAS Key laboratory for Research in Galaxies and Cosmology,
Department of Astronomy, University of Science and Technology of China, 
Hefei, 230026, China; jiazheng@mail.ustc.edu.cn, jnac@ustc.edu.cn}
\affiliation{School of Astronomy and Space Sciences,
University of Science and Technology of China, Hefei, 230026, China}

\author[0000-0003-4225-5442]{Yibo~Wang}
\affiliation{CAS Key laboratory for Research in Galaxies and Cosmology,
Department of Astronomy, University of Science and Technology of China, 
Hefei, 230026, China; jiazheng@mail.ustc.edu.cn, jnac@ustc.edu.cn}
\affiliation{School of Astronomy and Space Sciences,
University of Science and Technology of China, Hefei, 230026, China}

\author[0000-0003-4156-3793]{Jian-Guo Wang}
\affiliation{Yunnan Observatories, Chinese Academy of Sciences, Kunming 650011, PR China}
\affiliation{Key Laboratory for the Structure and Evolution of Celestial Objects, Yunnan Observatories, Kunming 650011, China}

\begin{abstract}
We report the discovery of a faint optical tidal disruption event (TDE) in the nearby star-forming galaxy NGC~3799. Identification of the TDE is based on its position at the galaxy nucleus, a light curve declining as $t^{-5/3}$, a blue continuum with an almost constant blackbody temperature of $\sim$12,000\,K, and broad ($\rm \approx15,000$\,km\,s$^{-1}$) Balmer lines and characteristic He~II 4686\,\AA\ emission. The light curve of AT~2023clx peaked at an absolute magnitude of $-17.16$\,mag in the $g$ band and a maximum blackbody bolometric luminosity of $4.56\times10^{42}$\,erg\,s$^{-1}$, making it the faintest TDE discovered to date. With a redshift of 0.01107 and a corresponding luminosity distance of 47.8\,Mpc, it is also the closest optical TDE ever discovered to our best knowledge. Furthermore, our analysis of {\it Swift}/XRT observations of AT~2023clx yields a very tight 3$\rm \sigma$ upper limit of 9.53$\rm \times10^{39}$\,erg\,s$^{-1}$ in the range 0.3--10\,keV. AT2023clx, together with very few other faint TDEs such as AT~2020wey, prove that there are probably a large number of faint TDEs yet to be discovered at higher redshifts, which is consistent with the prediction of luminosity functions (LFs). The upcoming deeper optical time-domain surveys, such as the Legacy Survey of Space and Time (LSST) and the Wide-Field Survey Telescope (WFST) will discover more TDEs at even lower luminosities, allowing for a more precise constraint of the low-end of the LF.
\end{abstract}


\section{Introduction}
A tidal disruption event (TDE) is the phenomenon observed
when a star comes too close to a supermassive black hole (SMBH). The star will be tidally disrupted and produce a radiation flare peaked at the ultraviolet (UV) to soft X-ray band, which usually happens in the core of the galaxy (\citealt{Rees1988,Gezari2021}). Although the first TDE was detected in X-ray (\citealt{Bade1996}), optical surveys have gradually dominated the discovery
of TDEs recently, especially since the operation of the Zwicky transient facility (\citealt{Bellm2019,Masci2019}). Moreover, a growing number of TDEs discovered in the infrared bands suggest that a considerable fraction of TDEs could be obscured by dust, which is missed by optical and X-ray surveys but can be revealed by their dust-reprocessed emission (\citealt{Mattila2018,Jiang2021}).

Recently, \citet{Yao2023} conducted a systematic analysis of demographics of TDEs using a sample of 33 optically selected TDEs from the ZTF survey over three years. They found an average peak of $<M_{g,peak}>=-19.91$\,mag for the $g$-band. However, there are a few nearby TDEs that are significantly fainter, such as iPTF16fnl at 70.8\,Mpc (\citealt{Blagorodnova2017,Brown2018,Onori2019}) and AT~2019qiz at 65.6\,Mpc (\citealt{Nicholl2020}), which belong to the spectroscopic class of H+He TDEs with Bowen fluorescence lines. \citet{Charalampopoulos2023} subsequently reported the faintest TDE in the ZTF sample at 119.7\,Mpc, AT~2020wey, and found that these three fast-decaying H+He TDEs lacked any other common properties. The diversity indicates that a large sample is needed to pin down the nature of these faint TDEs.
In reality, faint TDEs could constitute the largest population of all TDEs, and we may be biased toward finding bright nuclear flares due to flux-limited wide-field surveys (\citealt{Lin2022a,Charalampopoulos2023,Yao2023}).

In this letter, we present the discovery of the faintest and closest optical TDE so far in the nearby galaxy NGC~3799 at a redshift of 0.01107, which corresponds to a distance of 47.8\,Mpc. This event was initially detected by the All Sky Automated Survey for SuperNovae (ASAS-SN; \citealt{Shappee2014}), and it was suspected to be a TDE that could still be rising by \citet{Taguchi2023} on February 26, 2023. We describe our follow-up observations and data reduction in Section 2, followed by the analysis of the photometric properties of AT~2023clx and the identification of it as a robust TDE candidate combined with spectral characteristics in Section 3. Finally, we briefly discuss our results and draw conclusions in Sections 4 and 5. We assume a cosmology with $H_{0} =70$ km~s$^{-1}$~Mpc$^{-1}$, $\Omega_{m} = 0.3$, and $\Omega_{\Lambda} = 0.7$.

\section{Observations and Data}

\begin{figure*}[ht]
\figurenum{1}
\centering
\begin{minipage}{1.\textwidth}
\centering{\includegraphics[angle=0,width=1\textwidth]{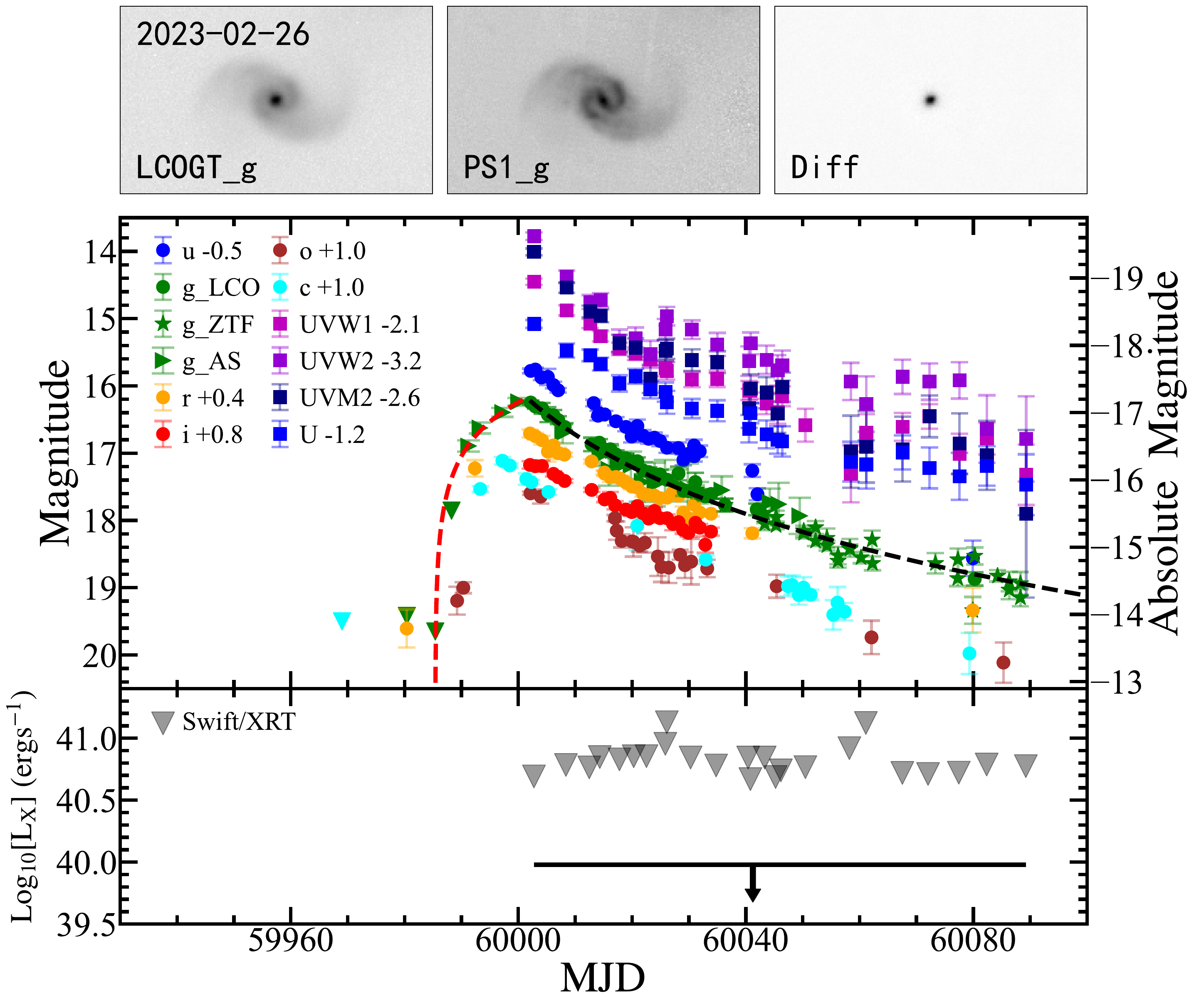}}
\end{minipage}
\caption{Top left: LCOGT g-band image which was taken on 2023-02-26 UT. Top middle: PanSTARRS g-band stack image of the host galaxy. Top right: Difference image of AT~2023clx. Middle: Host-galaxy-subtracted light curves of AT~2023clx in the UV and optical bands. The triangles represent the upper limits, and the dashed curve shows the power-law profile fit to the AT~2023clx g-band curve (red line for rise and black $t^{-5/3}$ line for decline). Bottom: the Swift/XRT light curve of AT~2023clx. The triangles represent the upper limits. AT~2023clx was non-detected in 0.3--10\,keV in any single observation, or even in the stacked image. The stacked X-ray upper limit is shown as the black horizontal line with arrows. (The data used to create this figure are available.)     \label{LCs}}
\end{figure*}

\subsection{Ground-based Optical Photometry}
We initiated optical $ugri$ band follow-up observations of AT~2023clx with the 1.0m Las Cumbres Observatory Global Telescope network (LCOGT; \citealt{LCOGT}) immediately after the event was reported on the Transient Name Server on 2023-02-26 UT (\citealt{Taguchi2023}). We used PanSTARRS (\citealt{PS1}) $gri$ band stack images as reference images and employed {\tt HOTPANTS}\,\citep{Becker2015} for image subtraction. Prior to subtraction, we removed cosmic rays and aligned the images using Astrometry.net. After image subtraction, we performed point spread function (PSF) photometry on the difference image using the Photutils package of Astropy \citep{Astropy} for the $gri$ data. For the $u$ band, we performed aperture photometry with a 5\arcsec \ aperture and then subtracted the host galaxy magnitude ($16.69\pm0.02$\,mag) with the same 5\arcsec \ aperture measured in the rescaled Sloan Digital Sky Survey (SDSS; \citealt{Gunn2006}) $u$ band image.

We tried to measure the position of AT~2023clx in the difference image of the LCO $r$-band image obtained on 2023-02-26 UT and the centroid of its host galaxy in the PanSTARRS $r$-band reference image by measuring the barycenter using the {\tt SExtractor}. The offset between them was measured to be $0.21\pm0.17$\,arcsec, taking into account the uncertainty during the alignment of the image. This corresponds to a physical offset of $49\pm40$\,pc at the distance of NGC~3799. Therefore, we can conclude that AT~2023clx is consistent with the center of the galaxy at the resolution level of the LCO images, making it a potential candidate for TDE.

We were intrigued by this event because we believed that it was still in the rising stage, given that its luminosity was 2 magnitudes fainter than the average peak luminosity of optical TDEs~\citep{Yao2023}. However, our initial two observations indicated that it was already in decline, suggesting that it might be a rarely-discovered faint TDE. In fact, its peak luminosity in the $g$ band ($M_g= -17.16$) is one of the faintest TDEs discovered thus far, comparable to that of iPTF16fnl (\citealt{Blagorodnova2017,Brown2018}; $M_g= -17.20$).

\subsection{Swift/UVOT photometry}
UV images were obtained with the {\it Neil Gehrels Swift Observatory} (hereafter {\it Swift}) with the Ultra-Violet Optical Telescope (UVOT). The {\it Swift} photometry (PIs: Gomez, Huang, Leloudas, and Wevers) were measured with UVOTSOURCE task in the {\tt Heasoft} package using 5\arcsec \ apertures after subtracting the galaxy background using {\it Swift}/UVOT images taken on July 18, 2010.
It was placed in the AB magnitude system \citep{OkeGunn83}, adopting the revised zero points and the sensitivity of \citet{Breeveld2011}.

\subsection{Swift/XRT photometry}
The X-ray Telescope (XRT) photometry was performed using XRTPIPELINE and XRTPRODUCTES. We used a circle with a radius of 47.1\arcsec\ as the source region and an annulus with an inner radius of 100\arcsec\ and an outer radius of 200\arcsec\ as the background. The source was X-ray faint in the observations, and we assumed an absorbed power-law spectrum with an index of $\Gamma=1.75$ \citep{ricci2017} and a Galactic hydrogen density of $2.51\times 10^{20}\,\text{cm}^{-2}$ \citep{HI4PI2016}. We then derived the 3$\sigma$ upper limit for the flux in the 0.3--10.0 keV range using WebPIMMS\footnote{\url{https://heasarc.gsfc.nasa.gov/cgi-bin/Tools/w3pimms/w3pimms.pl}}. In addition, we stacked the 26 event files ranging from MJD 60002.74 to 60089.23, deriving the upper limit of the mean luminosity of $9.53\times 10^{39}$ \lum\ in 3$\sigma$ level, with a total exposure time of 35.79 ks.

\subsection{Archival photometry Data}
We also collected host-subtracted light curves of AT~2023clx from public time domain surveys, including data from the Asteroid Terrestrial Impact Last Alert System (ATLAS; \citealt{Tonry2018,Smith2020}), the Zwicky Transient Facility (ZTF; \citealt{Masci2019}) and the All Sky Automated Survey for SuperNovae (ASAS-SN, \citealt{Shappee2014,Kochanek2017}).

The ATLAS $c-$ and $o-$band light curves were obtained using the ATLAS Forced Photometry Service, which produces PSF photometry on the difference images. ATLAS has $3-4$ single exposures within each epoch (typically within one day), so we binned the light curve every epoch to improve the signal-to-noise ratio (SNR). The ZTF light curves were obtained using the Lasair alert broker\footnote{Website link: https://lasair-ztf.lsst.ac.uk/} (\citealt{Smith2019}). The ASAS-SN host-subtracted g-band light curves were obtained using the ASAS-SN Sky Patrol photometry pipeline. We also collected and binned the photometry data of ASAS-SN. Considering the depth of g band of ASAS-SN is roughly 18.5\,mag (\citealt{deJ2022}), we only use the photometry data brighter than 18\,mag because the data fainter than 18\,mag have larger error than our LCO and ZTF photometry.


All light curves, after correction for Galactic extinction, are shown in Figure~\ref{LCs}. We assumed a \citet{Cardelli1989} extinction law with $R_V=3.1$ and a Galactic extinction of $E(B-V)=0.0268\pm0.0003$\,mag (\citealt{Schlafly2011}).

For better comparison of the peak $g$-band luminosity between AT~2023clx and the ZTF TDE samples, we tried to fit the profile of the light curve profile of AT~2023clx following the method of \citet{Yao2023} (see Section 3.2 for details). The rise function was poorly constrained with either Gaussian or power-law function due to only three detected points in ASAS-SN. Hence, we combined the upper limit to estimate the rise function. Shown in Figure \ref{LCs}, AT~2023clx can roughly be described by a rise in power law (index n = 0.64) and a $(t-t_0)^{-5/3}$ power-law decline ($t_0=\rm MJD\,59977\pm2$\,d) with a peak $g$-band luminosity of log$L_g=$\,42.31, which is the faintest TDE discovered by the ZTF survey to date compared to \citet{Yao2023}.

\subsection{Optical Spectra Observation and Data Reduction}
\begin{figure}
\figurenum{2}
\flushleft
\begin{minipage}{0.5\textwidth}
\flushleft{\includegraphics[angle=0,width=1\textwidth]{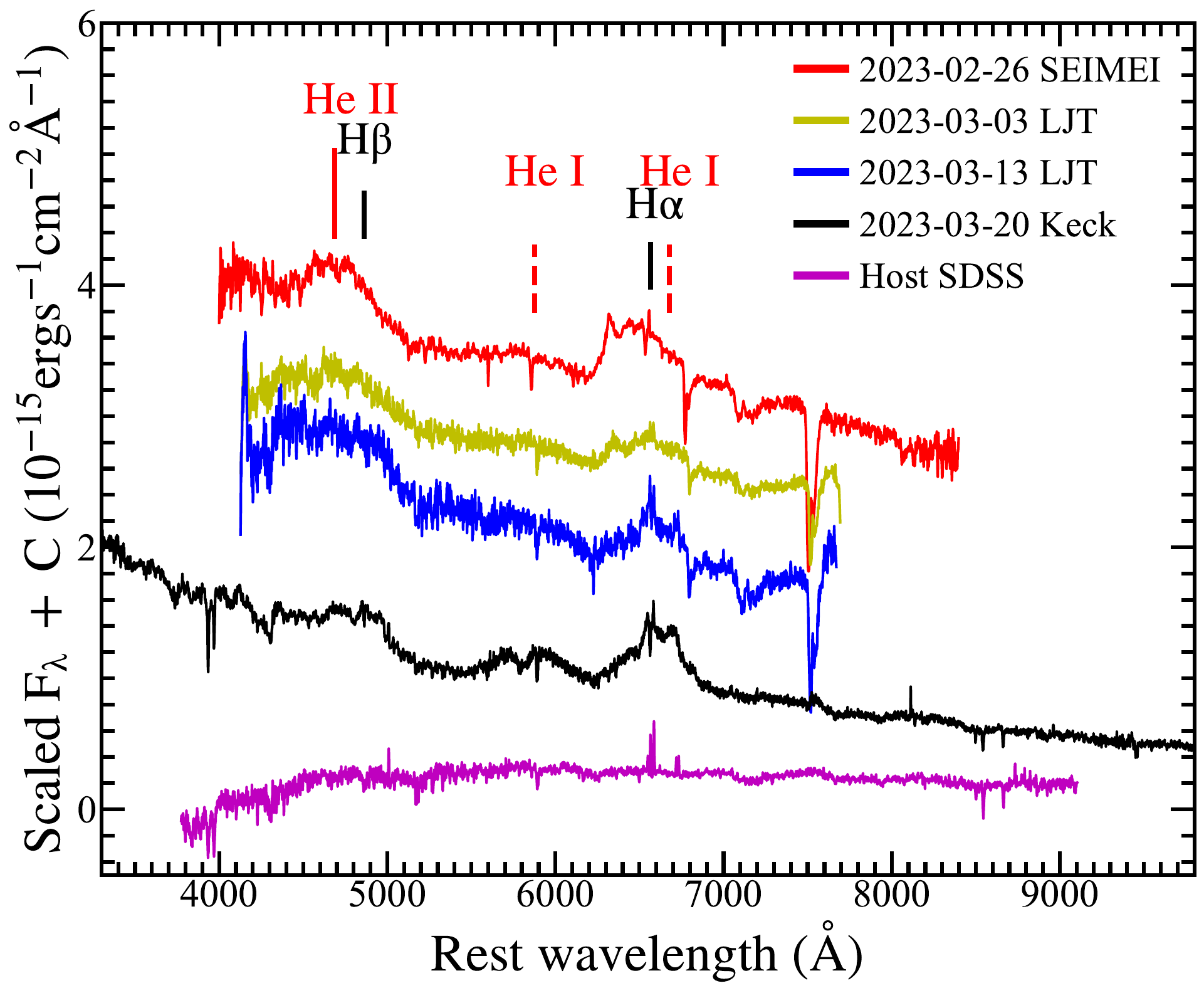}}
\end{minipage}
\caption{The spectra acquired for AT~2023clx. The identified H and He lines are labeled at the top (Red for  helium and black for hydrogen).
\label{spec_all}}
\end{figure}

Five spectra were acquired for AT~2023clx, including one from SDSS DR7, one observed by \citet{Taguchi2023} with the SEIMEI telescope\footnote{
We collected this spectrum from TNS website: https://www.wis-tns.org/object/2023clx} on 2023-02-26 UT, one observed by the ZTF group with the Keck telescope\footnote{
ZTF group upload this spectrum to another transient by mistake: https://www.wis-tns.org/object/2018meh} on 2023-03-20 UT and two obtained by ourselves with YFOSC onboard the LiJiang 2.4m telescope~(\citealt{Fan2015-2m4,Wang2019-2m4}) on 2023-03-03 UT and 2023-03-13 UT. 
We reduced the LJT spectra according to the standard procedure for long-slit spectra with PyRAF, a Python-based package of IRAF(\citealt{Tody1986,Tody1993}). All spectra are shown in Figure \ref{spec_all}.

\begin{figure}
\figurenum{3}
\flushright
\begin{minipage}{0.5\textwidth}
\flushright{\includegraphics[angle=0,width=1\textwidth]{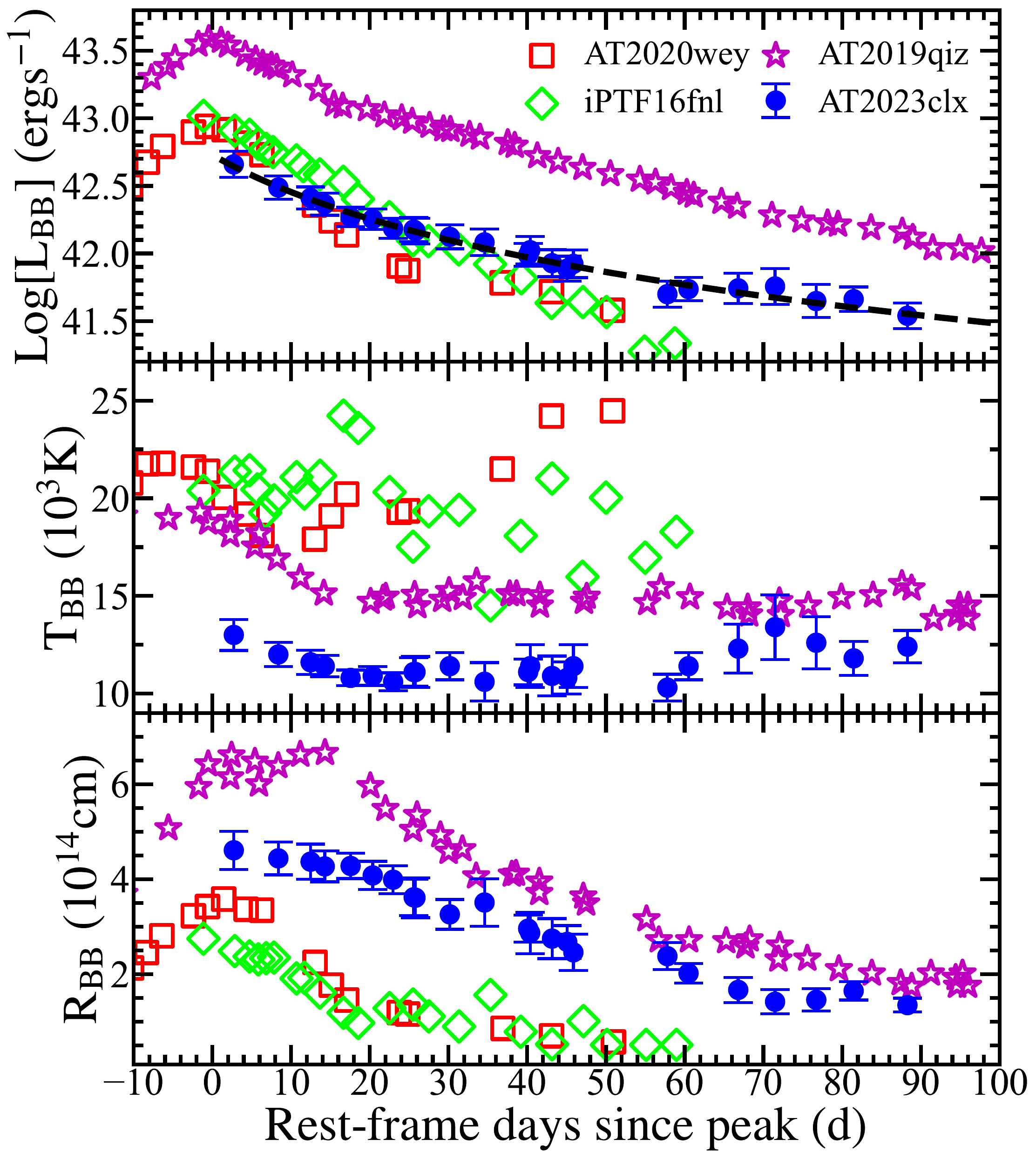}}
\end{minipage}
\caption{Evolution of the luminosity compared with iPTF16fnl (\citealt{Blagorodnova2017}), AT~2019qiz (\citealt{Nicholl2020}) and AT~2020wey (\citealt{Charalampopoulos2023}) {\it (top panel)}, temperature {\it (middle panel)}, and blackbody radius {\it (bottom panel)} of AT~2023clx. The dashed curve shows the $(t-t_0)^{-5/3}$ power-law ($t_0=\rm MJD\,59978\pm4$\,d) fit to the AT~2023clx bolometric light curves after the peak and the reference epoch is set to be the $g$-band peak at MJD = 60000.0.
\label{bbfit}}
\end{figure}

\begin{figure*}
\figurenum{4}
\centering
\begin{minipage}{1\textwidth}
\centering{\includegraphics[angle=0,width=1\textwidth]{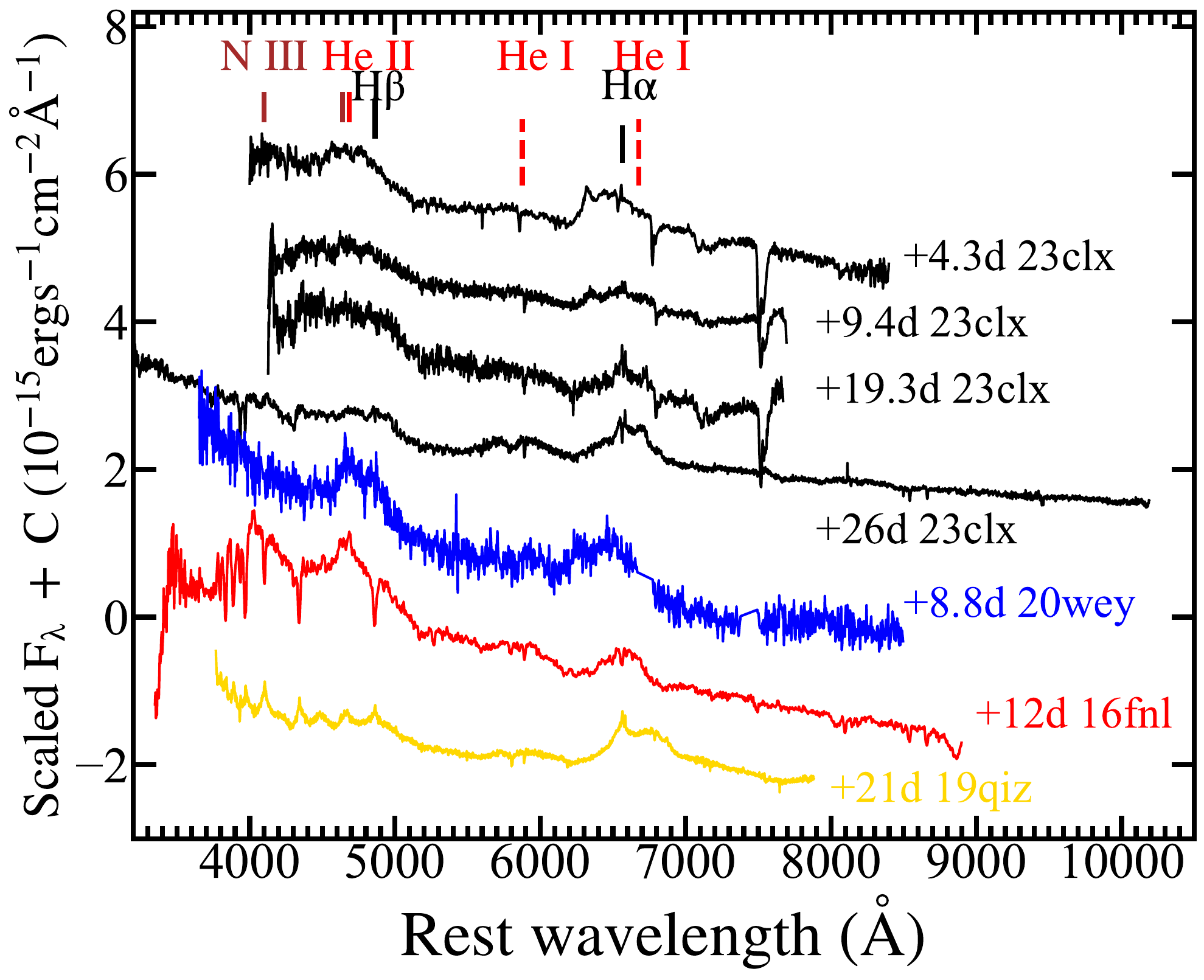}}
\end{minipage}
\caption{The comparison of optical spectra between AT~2023clx and some other low-luminosity TDEs at similar phases, including iPTF16fnl (\citealt{Blagorodnova2017}), AT~2019qiz (\citealt{Nicholl2020}) and AT~2020wey (\citealt{Charalampopoulos2023}). Their spectroscopic evolution is nearly identical.
\label{spec_compare}}
\end{figure*}

\section{Analysis and Results}

\subsection{Photometric and Spectral Analysis}
First, we use the package {\tt SUPERBOL} (\citealt{Nicholl2018})
to fit the spectral energy distribution (SED) of AT~2023clx with a blackbody model. The observing cadence after MJD~60040 in $r,\ i$ and $u$ bands is quite sparse. So we use third-order polynomials only to interpolate the $g,\ c$ and $o$ bands light curve in the late phase to match with the {\it Swift} epochs. The evolution of blackbody luminosity, temperature, and effective radius are shown in Figure \ref{bbfit}, compared with other faint TDEs. 

The optical/UV light curves of TDEs usually show a monthly rise followed by a decay with a timescale of months to years, sometimes following a $t^{-5/3}$ power-law decline (\citealt{Gezari2021,Velzen2021}). In the case of AT~2023clx, it exhibited a similar decline behavior that was consistent with a $t^{-5/3}$ power-law decline in both the $g$ band and the bolometric light curves (see Figures \ref{LCs} and \ref{bbfit}). The blackbody temperature was approximately 11,000--13,000\,K, which slowly decreased at the peak and was followed by a slow return to 13,000\,K. The blackbody radius $R_{BB}$ was approximately $4.61 \times 10^{14}$\,cm at the peak, followed by a decline. It is worth noting that AT2023clx has the lowest temperature among these faint TDEs, albeit with a medium blackbody radius. As a result, the maximum blackbody luminosity of AT~2023clx is only $(4.56\pm0.53) \times 10^{42}$\,erg\,s$^{-1}$, making it even fainter than previous low-luminosity TDEs (e.g., iPTF16fnl; \citealt{Blagorodnova2017}, AT~2019qiz; \citealt{Nicholl2020}, AT~2020wey; \citealt{Charalampopoulos2023}). Furthermore, these faint TDEs (e.g., AT~2020wey; \citealt{Charalampopoulos2023}) all exhibit fast declining light curves that are much steeper than the canonical $t^{-5/3}$. In the case of AT~2023clx, the decline appears to be less extreme than that of this subclass of TDEs. However, the limited number of samples at present cannot reveal their true statistical characteristics. 

We then compared the optical spectra of these low-luminosity TDEs in Figure~\ref{spec_compare}. Clearly, all these spectra near the peak display a blue continuum with broad $\rm H_{\alpha}$ and $\rm H_{\beta}$ emission, as well as the typical TDE emission line He~II 4686\,\AA. The broad $\rm H_{\alpha}$ is blended with He~I 6678\,\AA, and broad He~I 5876\,\AA\ line was also evident in the +26\,d Keck spectrum. The full width at half maximum (FWHM) of the broad Balmer lines is $\approx$15,000\,km\,s$^{-1}$. The $\rm H_{\alpha}$ profile is asymmetric, with a blueshifted peak at +4.3\,d and a redshifted peak at +26\,d (see Figure \ref{spec_compare}). A sharp peak on the blue side coincides in wavelength with [O~I] 6300\,\AA, which would indicate the presence of low ionization, low density, and slow-moving gas.
The red side may include some contribution from He~I 6678\,\AA, but this is likely not a major contributor, as we see only a very weak He~I 5876\,\AA, similar to AT~2019qiz (\citealt{Nicholl2020}). \citet{Nicholl2020} eventually attributed the evolution of such $\rm H_{\alpha}$ profiles of AT~2019qiz to an outflow. 
In addition, as we have seen the asymmetric broad $\rm H_{\alpha}$ profile, the N~III 4640\,\AA\ also could be blended with He~II and $\rm H_{\beta}$. Furthermore, most of the spectra in Figure \ref{spec_compare} do not cover the range of N~III 4100\,\AA, and it is hard to confirm whether the weak feature at +26\,d is real or not. Thus, the N~III 4100\,\AA\ can not be ruled out either.

However, the lack of low-ionization metal features (e.g., oxygen and calcium) in spectra near the peak almost excludes the typical Type-II supernovae scenario (\citealt{Filippenko97}). Type-IIn supernovae also show strong broad Balmer lines, but sometimes with high ionization lines of intermediate width that can be regarded as a result of CSM interaction (e.g., SN~2005ip, SN~2006jd and SN~2010jl; \citealt{SN2005ip&SN2006jd,SN2010jl}). If the CSM interaction of Type-IIn is relatively weak compared to the above three, the Fe~II and Ca~II P-Cygni lines appear again in their spectra (\citealt{Taddia2013}). Furthermore, the optical light curves of Type-IIn are relatively long-lived compared to AT~2023clx and normal supernovae do not stay at a temperature as high as 13,000\,K for months.

Therefore, AT~2023clx is consistent with a faint TDE scenario in both aspects of light curves and spectroscopic properties. Based on this, we classify AT~2023clx as a robust faint hydrogen and helium (H+He) optical TDE.

\subsection{Host-galaxy Properties}
NGC~3799, the host of AT~2023clx, is a well-resolved face-on spiral galaxy~(see top panels of Figure~\ref{LCs}). The pre-outburst SDSS spectra centered on the galaxy center have placed it in the locus of low-ionization nuclear emission-line region region (LINER) in the Baldwin-Phillips-Terlevich (BPT) diagram (\citealt{Baldwin1981,Veilleux1987}) based on its narrow-line ratios, indicating weak active galactic nuclear (AGN) activity albeit without broad emission lines.

\begin{figure}
\figurenum{5}
\flushleft
\begin{minipage}{0.5\textwidth}
\flushleft{\includegraphics[angle=0,width=1\textwidth]{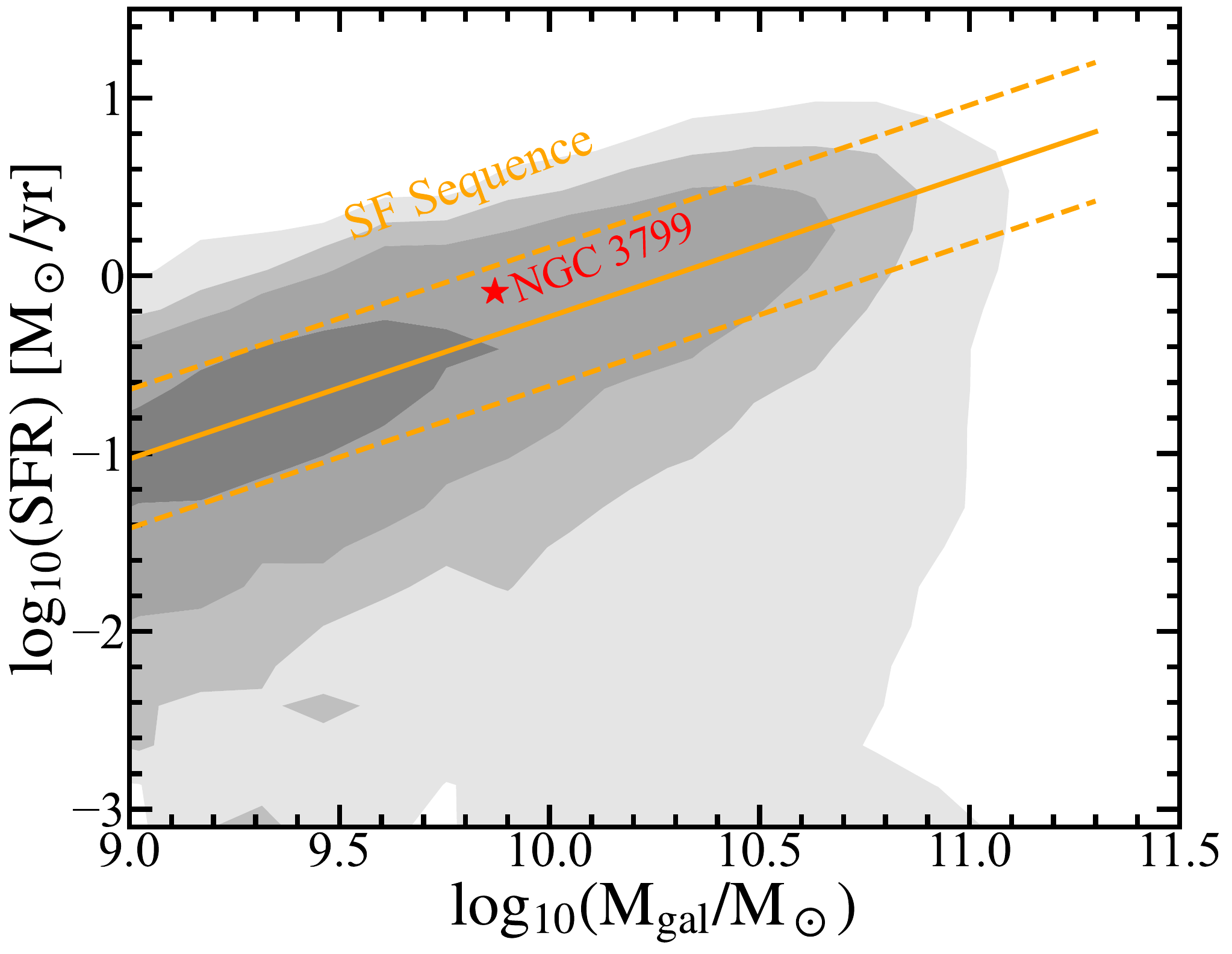}}
\end{minipage}
\caption{Distribution of galaxies in the SFR vs the $\rm M_{stellar}$ of galaxies from \citealt{Chang2015}. The star-forming sequence is shown in orange where the dashed lines show the 1$\sigma$ scatter. The gray scale represents the number of galaxies after weighting by 1/$\rm V_{max}$. The red star symbol marks the NGC~3799, which is well located on SF Sequence region.\label{SF_sequence}}
\end{figure}

\citet{Ramos2020} performed a spectral energy distribution (SED) analysis of a sample of 189 nearby galaxies, including NGC~3799. Their fitting considered the AGN emission and obtained a stellar mass of log$(M_*/M_{\odot}) = 9.87\pm0.02$ and a star formation rate (SFR) of $\rm logSFR = -0.09\pm0.02$ for NGC~3799. Consequently, it is located on the main sequence of star-forming galaxies (SF) (\citealt{Chang2015}; see Figure~\ref{SF_sequence}). AT~2023clx thus belongs to a TDE in a typical SF galaxy, whereas known optical TDEs show a preference for post-starburst (or green valley) galaxies (\citealt{Arcavi2014,French2016,French2020,Hammerstein2021}). Additionally, the fractional AGN contribution of NGC~3799 is indeed small ($f_{AGN}\textless0.2$) according to their fitting. We emphasize that the overall star formation activity does not conflict with its LINER classification in the BPT diagram since the central regions of the SF galaxies are commonly found to be quenched~(\citealt{Tacchella2015,Ellison2018}). Using the empirical relation between \mbh\ and the total galaxy stellar mass in the local universe (\citealt{Reines2015}): 
\begin{equation} 
\rm log(M_{BH}/M_{\odot})=\alpha + \beta\ log(M_{stellar}/10^{11}M_{\odot})
\end{equation} 
The central \mbh\ in NGC~3799 is estimated to be $10^{6.26\pm0.28}M_{\odot}$, using $\alpha=7.45\pm0.08$ and $\beta=1.05\pm0.11$.

\section{Discussion}
                                                                
Although only a handful of TDEs have been discovered to be faint ($M_g>-18$), the luminosity functions (LFs) of optical TDEs suggest that their real number should be large, as all of these LFs show a rising trend toward the low end (\citealt{Velzen2018,Lin2022b,Charalampopoulos2023,Yao2023}). However, an accurate LF profile requires more precise constraints, particularly when discovering more faint TDEs.

\begin{figure}
\figurenum{6}
\flushright
\begin{minipage}{0.5\textwidth}
\flushright{\includegraphics[angle=0,width=1\textwidth]{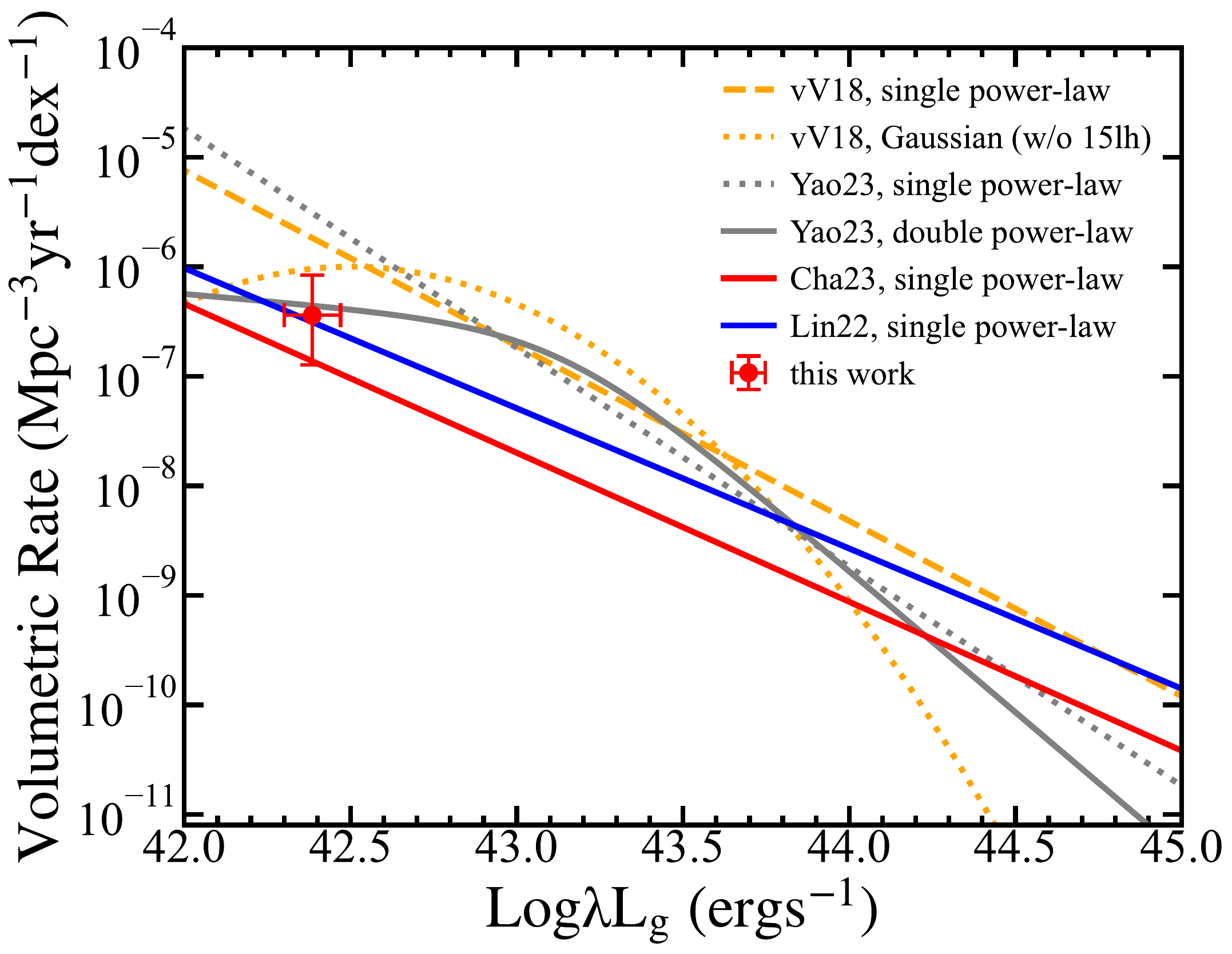}}
\end{minipage}
\caption{Our estimated volumetric rate using AT~2020wey and AT~2023clx is marked by red point, compare with the TDE luminosity function from \citet{Velzen2018,Lin2022b,Charalampopoulos2023} and \citet{Yao2023}.
\label{LF}}
\end{figure}

As one of the faintest TDEs discovered by the ZTF survey, AT~2023clx could provide the first data point at $L_{g}\textless10^{42.4}$\,erg\,s$^{-1}$. We also used the $\rm``1/\mathcal{V}{max}"$ method described in \citet{Lin2022b} to estimate the volumetric rate and considered the latest TDE selection criteria from the ZTF survey (\citealt{Yao2023}). The $\mathcal{V}{max}$ is defined as: \begin{equation} 
\mathcal{V}{max}\equiv V(z{max})~A_{survey}\times\tau_{survey} 
\end{equation} 
For the survey duration ($\tau_{survey}$), we set a starting date of 2018-10-1 UT and an end date of 2023-5-1 UT. Consistent with \citet{Lin2022b}, the effective survey area is set to $A\approx15000$, dg$^2$. We binned the two faintest ZTF TDEs (AT~2020wey and AT~2023clx) to estimate the volumetric rate in the faint end~($L_g\textless10^{42.5}$\,erg\,s$^{-1}$) and compared it with known optical TDE LFs~(see Figure \ref{LF}). We found that our result is about a factor of two higher than that measured by \citet{Charalampopoulos2023}, which is easily understood since they had only AT~2020wey when they constructed the LF. The data, although with a large uncertainty, seem to be well consistent with both the double power-law fitting measured by \citet{Yao2023} and the single power-law fitting measured by \cite{Lin2022b}. However, the volumetric rate at the low end is much lower than that of \citet{Velzen2018}. The general overestimation in \citet{Velzen2018} is probably due to sources that only have post-peak light curves in their sample.(\citealt{Lin2022b,Yao2023}).

Furthermore, we used the double power-law TDE luminosity functions measured by \citet{Yao2023}, which is consistent with our result, to estimate the proportion of faint TDE. Events that are fainter than or equally faint with AT~2019qiz constitute $\sim$74$\%$ in the currently observed $g$-band peak luminosity space ($L_g\sim 42.3-44.7$). This result is greater than the estimation obtained by \citet{Charalampopoulos2023}~(roughly 50--60$\%$). Again, we conclude that faint TDEs are not rare by nature, but constitute a large portion of the entire population, as mentioned by \citet{Charalampopoulos2023}.
The AT~2023clx might reveal that the contribution of the faint end may be even greater than previously thought.


It is worth noting that the distance used in this work is taken as a simple cosmological deduction from the redshift. The precise distance measurement for nearby galaxies is challenging, and thus previous works on faint nearby TDEs all adopted the same approach as us (e.g., 66.6\,Mpc for iPTF16fnl; \citealt{Blagorodnova2017}, 65.6\,Mpc for AT~2019qiz; \citealt{Nicholl2020}).
However, at these short distances, peculiar velocities and some redshift independent distances are not negligible, which might significantly affect the luminosity estimate of AT~2023clx significantly. \citet{Mould2000} performed a flow field correction on the nearby secondary distance indicators, including three attractors: the Local Supercluster, the Great Attractor and the Shapley Supercluster. The corrected recessional velocity of NGC~3799 in this way is $3823\pm31$\,km/s and the corrected distance is $54.6\pm4.6$\,Mpc, resulting in a peak $g$ band magnitude of $-17.45\pm0.12$\,mag and a peak blackbody luminosity of $\rm L_{bb}=(5.95\pm1.23) \times 10^{42}$\,erg\,s$^{-1}$ for AT2023~clx. Although the luminosity has increased by $\sim30\%$, it remains lower than iPTF16fnl ($1.0\pm0.15\times 10^{43}$\,erg\,s$^{-1}$; \citealt{Blagorodnova2017}) and AT~2020wey ($8.74\pm0.69\times 10^{42}$\,erg\,s$^{-1}$; \citealt{Charalampopoulos2023}).

We carefully examined other distance estimates for NGC3799 from the NASA/IPAC Extragalactic Database (NED), such as a 3K cosmic microwave background (CMB) correction distance of $52.3\pm3.8$\,Mpc and a Galactocentric (GSR) distance of $46.4\pm3.4$\, Mpc.  They are all less than the above value of $54.6\pm4.6$\,Mpc and thus lead to a lower correction of luminosity. Based on this, we confidently conclude that AT~2023clx is the faintest optical TDE observed to date.

\section{Conclusion}
In this work, we report the discovery of a new faint TDE in NGC~3799, a main-sequence star-forming galaxy located at a distance of only $\sim50$~Mpc. It holds the lowest peak blackbody luminosity and the closest distance among all optical TDEs discovered up to now. The main properties of AT~2023clx discovered by us are summarized below:

\begin{enumerate} 

\item[$\bullet$] The peak blackbody luminosity $\rm L_{bb}=(4.56\pm0.53) \times 10^{42}$\,erg\,s$^{-1}$ is lower than that of all other low-luminosity TDEs, although its absolute magnitude in the $g$ band ($M_{g}=-19.16$) is comparable to iPTF16fnl. It is also the closest optical TDE with a redshift of 0.01107 or a luminosity distance of 47.8~Mpc.

\item[$\bullet$] Both the optical/UV light curves and the bolometric light curve show a $t^{-5/3}$ power-law decay after the peak, which is not as fast as that of other faint TDEs discovered before.

\item[$\bullet$] AT~2023clx was not detectable in X-rays by {\it Swift}/XRT in any single observations or even in the stacked image. This yields a very tight 3$\rm \sigma$ upper limit of 9.53$\rm \times10^{39}$\,erg\,s$^{-1}$ in the range of 0.3--10\,keV.

\item[$\bullet$] The spectra taken around the optical peak show a strong blue continuum and broad Balmer lines blended with helium features (FWHM $\approx$15,000\,km\,s$^{-1}$), which is reminiscent of other faint TDEs. 

\end{enumerate}

AT~2023clx is the second optical TDE with a peak $g$ band luminosity of $L_g\textless10^{42.5}$\,erg\,s$^{-1}$ in the ZTF survey. The addition of AT~2023clx increases the volumetric rate at the extreme faint end by a factor of two compared to that given by \citet{Charalampopoulos2023}. This finding further demonstrates that the luminosity function (LF) is continuously rising towards the low end, and there are likely many more faint TDEs waiting to be discovered. The rarity of reported faint TDEs is simply due to selection bias, as we are biased towards finding bright events due to the flux-limited surveys, as mentioned by \citet{Charalampopoulos2023}. The upcoming deeper surveys, such as the Legacy Survey of Space and Time (LSST; \citealt{Ivezic2019}) and the Wide-Field Survey Telescope (WFST; \citealt{wfst2023,Lin2022a}), will undoubtedly find more faint TDEs, which will help us measure their rate and constrain the lower end of the TDE LF more precisely.  \\

We thank the anonymous referee for an extremely quick response and for providing valuable comments, which help to improve the manuscript. This work is supported by the SKA Fast Radio Burst and High-Energy Transients Project (2022SKA0130102), the National Natural Science Foundation of China (grants 11833007, 12073025, 12192221), and the 111 Project for "Observational and Theoretical Research on Dark Matter and Dark Energy" (B23042). We acknowledge the support of the Cyrus Chun Ying Tang Foundations. This research uses data obtained
through the Telescope Access Program (TAP), which has been
funded by the TAP member institutes. The authors acknowledge the support of the Lijiang 2.4m telescope staff. Funding for the telescope has been provided by the Chinese Academy of Sciences and the People's Government of Yunnan Province”. The ZTF forced photometry service was funded under the Heising-Simons Foundation grant \#12540303 (PI: Graham).

\end{document}